\documentclass[twoside,fleqn]{ActaStyle}
\usepackage{times}

\usepackage{lscape}
\usepackage{graphicx,epsfig}
\usepackage[tbtags]{amsmath}

\newcommand{\dd}{\mathrm{d}}

\newcommand{\bea}{\begin{eqnarray}}   
\newcommand{\eea}{\end{eqnarray}}
\newcommand{\demu}{\partial_{\mu}}

\newcommand{\de}{\partial}
\newcommand{\bear}{\begin{array}{l}}
\newcommand{\eear}{\end{array}}

\newcommand{\ldl}{\Lambda \partial_{\Lambda}}
\newcommand{\inte}{\! \int \!\!}
\newcommand{\ie}{{\it i.e.}\ }
\newcommand{\eg}{{\it e.g.}\ }
\newcommand{\etal}{{\it et al.}}

\newcommand{\etc}{{\it etc.}\ }

\newcommand{\one}{\hbox{1\kern-2mm l}}

\def\lam{\lambda}
\def\Lam{\Lambda}
\def\de{\partial}

\def\0{\vec{0}}

\def\sig3{\sigma_3}

\def\half{{\textstyle{1\over2}}} 
\def\ker#1{\cdot #1 \cdot}

\def\e{{\rm e}}

\def\hS{\hat{S}}

\def\one{\hbox{1\kern-.8mm l}}

\def\eq#1{eq.~(\ref{#1})}

\def\s#1#2#3{S^{(#1)}_{#2} (#3)}
\def\phi{\varphi}

 \def\hepth#1{{\tt hep-th/#1}}

\def\authorheading{S.~Arnone et al.}

\def\shorttitle{Exact scheme independence}

\begin{document}

\pagerange{1}{6}

\title{%
A DEMONSTRATION OF SCHEME INDEPENDENCE IN SCALAR ERGs
}

\author{
Stefano Arnone\email{S.Arnone@soton.ac.uk}, Antonio
Gatti\email{A.Gatti@soton.ac.uk}, Tim R.~Morris\email{T.R.Morris@soton.ac.uk}
}
{
Department of Physics and Astronomy, 
University of Southampton\\
Highfield, Southampton SO17 1BJ, 
United Kingdom}

\day{May 10, 2002}

\abstract{%
The standard demand for the quantum partition function to be invariant under 
 the renormalization group transformation results in a general class of  
exact renormalization group equations, 
 different in the form of the kernel. Physical quantities should not be 
 sensitive to the particular choice of the kernel. Such scheme independence
is elegantly illustrated in the scalar case by showing that, even with a
general kernel, the 1-loop beta function may be expressed only in terms of
the effective action vertices, and in this way the universal result is 
recovered.}

\section{Introduction}
\label{sec:intr} \setcounter{section}{1}\setcounter{equation}{0}
The non-perturbative meaning of renormalization, as understood
by Wilson, is formulated most directly in the continuum in terms
of the Exact Renormalization Group (ERG)~\cite{Wil}. Moreover, the
fact that solutions $S$ of the corresponding flow equations
can then be found directly in terms of renormalized quantities, that
all physics (\eg Green functions) can be extracted from this
Wilsonian effective action $S$, and that renormalizability is 
trivially preserved in almost any approximation \cite{morig}, turns these 
ideas into a powerful framework for considering non-perturbative
analytic approximations \cite{Wil, morig}, \cite{We}--\cite{LM}.

Recently, very general versions
of the ERG \cite{red}
have been considered, dependent on the choice of
a functional $\Psi$, known as the ``kernel'' of the ERG \cite{Mo1,LM}.
In particular, each ERG
is associated with a $\Psi$, that induces a
reparametrisation (field redefinition) along the flow, and acts
as a connection between the theory space of actions at different
scales $\Lambda$. As a result, local to some generic point $\Lambda$
on the flow, all the ERGs, including these generalised ones,
may be shown to be just reparametrisations of each other. 
When this reparametrisation can be extended globally, the result is
an immediate proof of scheme
independence for physical observables. Indeed computations of physical
quantities then differ only through some field reparametrisation.
One practical example
is an explicit field redefinition that interpolates between 
results computed using different choices of cutoff function 
$c(p^2/\Lambda^2)$ \cite{LM}.

Obstructions to full (global) equivalence of ERGs can arise, 
on the one hand from differences in the global structure of fixed points 
deduced from the two flows,
and on the other from the non-existence at special points of an
inverse in an implied change of variables (from $S$ to $\Lambda$) 
\cite{LM}. However it is difficult to  
determine these, given that 
in practice one has to make approximations 
in order to solve these theories. Furthermore, computations within a 
generalised ERG, such as the type being used for gauge theory 
\cite{Mo1,Mo},
generate many more terms, whose interpretation
seems obscure \cite{Mo1,Mo}. 

This contribution (summarising \cite{scalar}) 
addresses these problems within a sufficiently simple
and bounded context: the one-loop $\beta$ function of massless four dimensional 
one component $\lambda\varphi^4$
scalar field theory. We will see that even for a very
general form of $\Psi$ (one involving a general `seed' action $\hS$),
the correct universal result is obtained. To our knowledge, this is the
first concrete
test of such scheme independence beyond testing for cutoff function
independence. The only requirements we have
to impose to recover the universal result,
are some very weak and general requirements which are
necessary in any case to ensure that the
Wilsonian action $S$ makes sense. To this level then, all such ERGs
are equivalent and merely parametrise changes of scheme. 

\section{From the Polchinski equation to general ERGs}
\label{sec:form}

The Polchinski equation~\cite{Po} for a self-interacting real scalar field
in four Euclidean dimensions takes the form
\be \label{sintfleq}
\ldl S^{int} = -{1\over\Lambda^2}
 {\delta S\over\delta\phi}^{int} \!\!\!\!\!\ker{c'}  
{\delta S\over\delta\phi}^{int}\!\!\! +{1\over\Lambda^2}{\delta 
\over\delta\phi} \ker{c'} {\delta S\over\delta\phi}^{int}\!\!\!\!\!\!,
\ee
where $S^{int}$ is the interaction part of the effective action. 
Prime denotes differentiation 
with respect to the function's argument (here $p^2/\Lambda^2$) and the 
following shorthand has 
been introduced: for two functions $f(x)$ and $g(y)$ and a 
momentum space kernel $W(p^2/\Lambda^2)$, 
with $\Lam$ being the effective cutoff, 
\be
f \ker{W} g =
\int\!\!\!\!\!\int\!\!\dd^4\!x\,\dd^4\!y\
f(x)\, W_{x y}\,g(y), \qquad W_{x y} = \inte {\dd^4 p \over (2\pi)^4} \,
W(p^2/\Lambda^2) {\rm e}^{i p \cdot (x-y)}.
\ee
$c(p^2/\Lambda^2)>0$ is a {\it smooth}, \ie infinitely
differentiable, ultra-violet cutoff profile, which modifies propagators $1/p^2$
to $c/p^2$. It satisfies $c(0)=1$ so that low energies are unaltered,
and $c(p^2/\Lambda^2)\to0$ as $p^2/\Lambda^2\to\infty$
sufficiently fast that all Feynman diagrams are ultra-violet regulated.

Note that the regularised kinetic term in the effective action may be
written as $\half\,\demu \varphi \ker{c^{-1}} \demu\varphi$.
This will be referred to as the seed action and denoted 
by $\hS$. In terms of the total effective action, $S=\hS+S^{int}$, 
and $\Sigma \doteq S-2\hS$, the ERG equation reads (up to a vacuum
energy term that was discarded in~\cite{Po}) 
\be \label{pofleq}
\ldl S = -{1\over\Lambda^2}
 {\delta S\over\delta\varphi} \ker{c'}  
{\delta \Sigma \over\delta\varphi} +{1\over\Lambda^2}{\delta 
\over\delta\varphi} \ker{c'} {\delta \Sigma \over\delta\varphi}
\,\, \Rightarrow \,\, \ldl \e^{-S} =
-{1\over\Lambda^2}{\delta\over\delta\varphi}\cdot c' \cdot \left(
{\delta\Sigma \over\delta\varphi}\,\e^{-S}\right),
\ee
\ie the infinitesimal RG transformation results in a change in the
integrand which is a total functional derivative.

Incidentally, this establishes a rather counterintuitive result, that
integrating out degrees of freedom is just equivalent to redefining the
fields in the theory \cite{Mo,LM}. In the present case, the change in
the partition function may be shown to correspond to the change of
variables $\varphi \rightarrow \varphi + \delta \Lam \, \Psi$, with the
``kernel'' $\Psi = -{1\over \Lam^3} \, c' \, {\delta \Sigma
\over\delta\varphi}$ that appears in \eq{pofleq} \cite{LM}.

Different forms of ERG equations correspond to choosing different
kernels $\Psi$. There is a tremendous amount of freedom in this choice, 
just as there is a great deal of freedom in choosing the form
of a blocking transformation in the condensed matter or lattice 
realisation of the Wilsonian RG \cite{Wil}.
The flow equation (\ref{sintfleq}) is distinguished only by its relative
simplicity (related to incorporating the cutoff only in the kinetic
term). Nevertheless, physical quantities should turn out to be universal
\ie independent of these choices.

As shown above, Polchinski's equation comes from setting the seed action
equal to the effective kinetic term in the Wilsonian effective action. If 
we are to reproduce that very term at the classical level,
the bilinear term in $\hS$
must continue to be equal to $\half\,\demu \varphi \ker{c^{-1}}
\demu\varphi$.
Furthermore, we choose to leave the $\phi\leftrightarrow-\phi$ symmetry 
alone, by requiring that $\hS$ is even under this symmetry. We are left 
with a generalised ERG parametrised by the 
infinite set of seed action $n$-point vertices, $n=4,6,8,\cdots$. We will
leave each of these vertices as completely unspecified functions of
their momenta except for the requirement that {\it the vertices be infinitely 
differentiable and lead to convergent momentum integrals}.
(The first condition ensures that no spurious infrared 
singularities are introduced and that all effective vertices can be
Taylor expanded in their momenta to any order \cite{Mo1,morig}.
The second is necessary for the flow equation to make sense
at the quantum level and also ensures the flow actually corresponds
to integrating out modes \cite{Mo,LM}.)

We are therefore incorporating in the momentum dependence of
{\sl each} of the seed action $n$-point vertices, $n=2,4,6,\cdots$, 
an infinite number of parameters. Remarkably, however, we can still compute
the one-loop $\beta$ function. Moreover, as we will see in the next
section, we can 
invert the flow equation by expressing $\hS$ vertices in terms of
$S$, and in this way demonstrate explicitly that the result is universal
- {\it viz.} independent of the choice of $c$ and $\hS$.
\section{One-loop beta function with general seed action}
\label{sec:oneloop}

As usual, universal results are obtained only after the imposition of 
appropriate renormalization conditions which allow us to define what we
mean by the physical (more generally renormalized) coupling and field. 
(The renormalized mass must also be defined and is here set to zero 
implicitly by ensuring that the only
scale that appears is $\Lambda$.) 

We write the vertices of $S$ as
$\s{2n}{}{\vec{p};\Lam}\equiv\s{2n}{}{p_1,p_2,\cdots,p_{2n};\Lam}$
(and similarly for the vertices of $\hS$). In common with earlier works 
\cite{Po,Bo}, we define the renormalized four point coupling $\lambda$ by 
the effective action's four-point vertex evaluated at zero momenta:
$\lambda(\Lambda)=\s{4}{}{\vec{0};\Lam}$. This makes sense once we 
express quantities in terms of the renormalized field,
defined (as usual) to bring the kinetic term
into canonical form. 

We will also rescale the field  $\varphi\mapsto
\frac{1}{\sqrt{\lambda}}\, {\varphi}$, 
so as to put the coupling constant in
front of the action. This ensures the expansion in the coupling constant 
coincides with the one in $\hbar$, which is more elegant, 
although it will introduce a `fake'
contribution to the anomalous dimension $\gamma$ in these variables.
It is also
analogous to the treatment pursued for gauge theory in refs. \cite{Mo1,Mo}
(where gauge invariance introduces further simplifications in particular
forcing $\gamma=0$ for the new gauge field). The following analysis thus
furnishes a demonstration that these ideas also work within scalar
field theory.  

Expanding the action, the beta function $\beta(\Lam) = \ldl \lam$ 
and the anomalous dimension in powers of the coupling constant yields 
the loopwise expansion of the flow equation
\begin{eqnarray}
\Lam\de_{\Lam}S_0=-\frac{1}{\Lam^2}\frac{\delta S_0}{\delta\varphi}\cdot
c'\cdot\frac{\delta (S_0-2\hat{S})}{\delta\varphi}\label{scalartree},
\phantom{\Lam\de_{\Lam}S_1-\beta_1 S_0-{\gamma_1\over2}\ 
\phi\!\cdot\!{\delta S_0\over\delta\phi}
=-\frac{2}{\Lam^2}\frac{\delta S_1}{\delta\varphi}\cdot
c'\cdot\cdot}\\
\Lam\de_{\Lam}S_1-\beta_1 S_0-{\gamma_1\over2}\ 
\phi\!\cdot\!{\delta S_0\over\delta\phi}
-\frac{2}{\Lam^2}\frac{\delta S_1}{\delta\varphi}\cdot
c'\cdot\frac{\delta(S_0-\hat{S})}{\delta\varphi}+\frac{1}{\Lam^2}\frac{
\delta}{\delta\varphi}\cdot
c'\cdot\frac{\delta(S_0-2\hat{S})}{\delta\varphi}\label{scalar1loop},
\\ \nonumber
\eea
\etc \hspace{-.4em}, where $S_0$ ($S_1$) is the classical (one-loop) effective action and
$\beta_1$ and $\gamma_1$ are the one-loop contributions to $\beta$ and
$\gamma$. These latter may now be extracted directly from 
\eq{scalar1loop},  as specialised to the two-point and four-point 
effective couplings, $\s{2}{}{\vec{p};\Lam}$ and $\s{4}{}{\vec{p};\Lam}$ 
respectively, once the renormalization conditions have been taken
into account. In these variables they read
\be
\label{r}
\s{2}{}{p,-p;\Lam} = \s{2}{}{0,0;\Lam} + p^2 +O(p^4/\Lam^2), \qquad
\qquad \s{4}{}{\vec{0};\Lam} = 1.
\ee
As usual, both conditions are  saturated at tree level.
Hence there must be no quantum corrections to the four-point 
vertex at $\vec{p} =
\vec{0}$, or to the $O(p^2)$ part of the two-point vertex.

The flow equations for these special parts of the quantum corrections
thus greatly simplify, 
reducing to algebraic equations which then determine the $\beta_i$
and $\gamma_i$. At one loop:
\bea
\beta_1+2\gamma_1&=&\frac{8c'_0}{\Lam^2}\Big[1-\hat{S}^{(4)}(\vec{0})\Big] 
\s{2}{1}{0}-\frac{1}{\Lam^2}\int_q c'({\textstyle {q^2 \over
\Lam^2}})\Big[S^{(6)}_0-2\hat{S}^{(6)}\Big](\vec{0},q,-q), \label{beta1}\\ 
\beta_1+\gamma_1&=&- \frac{1}{\Lam^2}\int_q c'({\textstyle {q^2 \over
\Lam^2}})
\Big[ \left.S^{(4)}_0-2 \hS^{(4)}\Big] (p,-p,q,-q)\right|_{p^2},\label{gamma1}
\eea 
where $c'_0 = c'(0)$, $\int_q \doteq \int {\dd^4 q \over (2 \pi)^4}$,
and 
the notation $|_{p^2}$ means that one should take the coefficient 
of $p^2$ in the series expansion in $p$. 

In order to evaluate \eq{beta1}, we need to
calculate $\s{2}{1}{0}$ and $\s{6}{0}{\vec{0}, q, -q}$. We would also need
$\hS^{(4)}(\vec{0})$ and $\hS^{(6)}(\vec{0},q,-q)$, but we 
can avoid using explicit expressions for them, and thus keep 
$\hS$ general, by using the equations of motion. 

Our general strategy (see also \cite{stefano}) is to use the equations
of motion to express any hatted vertices in terms of the effective
ones. This will cause almost all the non-universal terms, those depending
on the details of the cutoff function and/or on the explicit form of the
seed action, to cancel out. The remaining ones will disappear once
$\gamma_1$ is substituted using \eq{gamma1}. The simplest way to appreciate
that non-universal terms cancel is to recognise they can be paired up into 
total $\Lam$-derivatives, which can be taken outside the momentum integral as
the integrand is regulated both in the ultra-violet and the infrared. 
Furthermore, as those
terms contribute to $\beta_1$, they must be dimensionless and thus cannot
depend upon $\Lambda$ after the momentum integral has been carried out,
hence the result vanishes identically! 

As an example, we use \eq{scalartree} to express $\hS^{(6)}(\0, q, -q)$ in 
terms of the effective action vertices,
\begin{eqnarray}
\label{sh6}
\hat{S}^{(6)}(\0,q,-q) = \frac{\Lam^2}{4q^2} \frac{c_q}{c'_q} \left\{ \ldl
\s{6}{0}{\0,q,-q} +\frac{8 c'_0}{\Lam^2}\Big [1-\hat{S}^{(4)}(\0) \Big]
\s{4}{0}{0,0,q,-q} \right.\nonumber\\
-2c'_0 \frac{c_q}{q^2c'_q} \ldl \s{4}{0}{0,0,q,-q}
\left.-\frac{6}{q^2} \s{4}{0}{0,0,q,-q} \ldl \Big[ c_q \,
\s{4}{0}{0,0,q,-q} \Big] \right\},
\end{eqnarray}
and substitute it back into \eq{beta1}. The first term in \eq{sh6} and the
$S_0^{(6)}$ term in \eq{beta1} pair up into 
$
\ldl \int_q \frac{1}{2 q^2} \, c({\textstyle
{q^2 \over \Lam^2}}) \, \s{6}{0}{\0,q,-q},
$
which vanishes since the result of a convergent dimensionless integral cannot
depend upon $\Lam$. Also, the second term in \eq{sh6} exactly cancels the
first term in \eq{beta1} once the one-loop two-point vertex at null
momentum is computed. This latter calculation is carried out exactly
the same way, namely by eliminating the four-point seed action vertex in
favour of the effective one. The result takes the form
\be
\s{2}{1}{0} = -\int_q \frac{c_q \, \s{4}{0}{0,0,q,-q}}{2 q^2},
\ee
with no integration constant since for a massless theory, 
there must be no other explicit scale in the theory apart from the 
effective cutoff.

Of the two remaining terms in \eq{beta1}, one is cancelled by $\gamma_1$,
while the other reads 
\be
\beta_1 = \frac{3}{2} \int_q \frac{1}{q^4}\, \ldl\, \left\{ c({\textstyle
{q^2 \over \Lam^2}}) \s{4}{0}{0,0,q,-q} \right\}^2, 
\ee
which is nothing but the
standard one-loop diagrammatic result for the $\beta$ function, written in
terms of the regularised propagator and the effective $S^{(4)}_0$.
Here the derivative with respect to $\Lam$ cannot be taken outside the
integral, as the 
latter would not then be properly regulated in the infrared. 
Indeed, it yields the standard one-loop result:\footnote{The term in braces 
depends only on $q^2/\Lam^2$. $\Omega_4$
is the four dimensional solid angle. The result follows from the
convergence of the integral and normalisation conditions $c(0)=1$ 
and \eq{r}.}
\be
\beta_1 = -\frac{3}{2}{\Omega_4\over(2\pi)^4}  
\int_0^{\infty}\!\!\! \dd q \, \de_q \left\{ c ({\textstyle
{q^2 \over \Lam^2}}) 
\, \s{4}{0}{0,0,q,-q} \right\}^2 = \frac{3}{16\pi^2}.
\ee
\section{Summary and conclusions}
\label{sec:sum}
Starting with the generalised ERG flow equation,
we computed tree level two, four and six point vertices. At one-loop
we computed 
the effective mass $\s{2}{1}{0}$ and wavefunction renormalization 
$\gamma_1$. By combining all these with the flow of the one-loop 
four-point vertex at zero momenta, we arrived at the expected universal 
result $\beta_1=3/(4\pi)^2$.

The flow equation we used 
differs from the Polchinski flow equation
(\ref{sintfleq}), equivalently \eq{pofleq}, because the seed action
$\hS$ is no longer set to be just the kinetic term, but
is generalised to include all arbitrary even higher-point
vertices. (We also scaled out the coupling $\lambda$.)

We then proceeded to compute the tree and one-loop 
corrections
exploiting the ability, within the ERG, to derive 
directly the renormalized couplings and vertices (\ie without having to
refer back to an overall cutoff and bare action).

We could now argue that we should have
expected these results, without the detailed calculation. 
Nevertheless this is the first specific test of these
ideas beyond that of just cutoff function independence, and in the
process we found the
restrictions on $\hS$ sufficient to ensure scheme independence
at this level. They are merely that the seed vertices be infinitely 
differentiable and lead to convergent momentum integrals, which
as we noted are necessary conditions in any case.

It is important to stress that many of our specific choices
(what we chose to generalise in $\Psi$, how we
incorporated wavefunction renormalization, organised and solved the 
perturbative expansion) are not crucial to the calculation. 
On the contrary there are very
many ways to organise the computation; we just chose our favourite
one. The crucial step in navigating the generalised corrections, 
appears to be the recognition
that one should eliminate the elements put in by hand, in this
case vertices of $\hS$, in favour of the induced solution, the Wilsonian
effective action $S$, which encodes the actual physics. 

For us, this is the
most important conclusion of the present paper since it implies a 
practical prescription for streamlined calculations which can be used 
even in more involved settings such as in the manifestly
gauge invariant framework
\cite{Mo1,Mo,ant}, where there is no equivalent calculation
one can directly compare to.

\begin{ack}
The authors wish to thank the organisers of the fifth international
Conference ``RG 2002'' for providing such a stimulating environment. 
T.R.M. and S.A. acknowledge financial support from PPARC Rolling Grant
PPA/G/O/2000/00464.   
\end{ack}

\end{document}